\documentclass[12pt]{article}
\usepackage[english]{babel}
\usepackage{amsfonts}
\usepackage{cite}
\usepackage{amsmath} 
\usepackage{graphicx}

\begin{document}

\begin{center}
{\large \bf Optimization Method in 2D Thermal Cloaking Problems}

{\bf G. V. Alekseev}$^{1,2}$ {\bf and  D. A. Tereshko}$^{1}$\\
\medskip
{\small $^{1}$Institute of Applied Mathematics FEB RAS, Vladivostok, Russia\\
$^{2}$School of Natural Sciences, Far Eastern Federal University, Vladivostok, Russia}
\end{center}

\begin{abstract}
Inverse problems associated with designing cylindrical 
thermal cloaking shells are studied.
Using the optimization method these inverse problems
are reduced to corresponding control problems in which 
the diagonal components of diagonal in polar coordinates 
tensor of thermal conductivities play the role of 
passive controls. A numerical algorithm based on the 
particle swarm optimization is proposed and the results 
of numerical experiments are discussed.
Rigorous optimization analysis shows that high cloaking efficiency of the shell 
can be achieved either using a highly anisotropic single-layer shell 
or using a multilayer shell composed of only three isotropic natural materials with optimally selected thermal conductivities.
\end{abstract}

\section{Introduction}
In recent years much attention has been given to creation of tools of material objects masking from
detection with the help of electromagnetic or acoustic location. Beginning with pioneering papers
\cite{Do61,Leo06,Pendry06} the large number of publications was devoted to developing different methods of solving
the cloaking problems. Transformation optics (TO) is the most popular method of designing cloaking
devices (hereafter, cloaks). The methodology of cloaking based on this method obtained the name
of direct design because it is based in fact on solving direct problems of electromagnetic scattering.

The first works in this field focused on the electromagnetic cloaking, i.e.
cloaking objects from detection by electromagnetic location. Then the main results of the electromagnetic cloaking
theory were expanded to an acoustic cloaking \cite{Cum07,Chen07} and to
cloaking magnetic, electric, thermal and other static fields \cite{Wood07,Chen08,Fan08,
San11,Go12, Guen12, Nara12, Dede13, Nara13, Gao13, Guen13,Leo13,Pet14,Alu14,Hu14, Sch14,HanAM14,HanPRL14,Xu14,Yang12,Han16,Han13,Han14, Hu15, Hu16, Raza16,Wang18}. 
It should be noted that the invisibility devices (hereafter, cloaks) designed on the
basis of direct strategies possess serious drawbacks. The main one is the difficulty of their
technical realization. For example, the design of the TO-based cloaks involves extreme values of
constitutive parameters and spatially varying distributions of the permittivity and permeability
tensors which are very difficult to implement \cite{Xu13}.

That is why the another cloak design strategy began develop recently. It obtained the name of
inverse design as it is related with solving inverse electromagnetic (acoustic or static) problems
(see \cite{Xu13,Mono16}). The optimization method forms the core of the inverse design methodology.
Just this enables to overcome some substantial limitations of previous cloaking solutions. 
A growing
number of papers is devoted to applying the inverse design methodology in various cloaking
problems. Among them we mention the fist papers  \cite{Xi09,Popa09} where numerical optimization algorithms are
applied for finding the unknown material parameters of TO-based cloak and 
papers \cite{Dede2014,Fuji2018, PeFa17,Fach2018} associated with the use of topology optimization and discrete material 
optimization for solving problems of designing cloaks, concentrators and other thermal functional devices.
Papers \cite{Al13c,AlLe16,Al18} are devoted to theoretical analysis of cloaking problems using the optimization approach.

Optimization method is applied and in this paper for solving inverse problems for the 2D stationary  model of heat transfer. 
These problems arise when developing the technologies of designing thermal cloaking devices  having the form of the cylindrical shell. It is assumed that the desired cloaking shell consists of finite number of  layers every of which is filled with homogeneous anisotropic medium. Radial and azimuthal thermal conductivities $k_r$ and $k_\theta$ of the medium filling every layer play the role of controls. As a result our cloaking problems are reduced to solving the corresponding finite-dimensional extremum problems. We propose a numerical algorithm of solving these finite-dimensional extremum problems which is based on using the particle swam optimization (PSO) (see \cite{Po07}) 
according to the scheme proposed in \cite{AlLeTe17c,AlLeTe17s} and discuss some results of numerical experiments.

\section{Statement of direct conductivity problem. Properties of exact solution} 
We begin with statement of the general direct conductivity problem, considered in a rectangle $D=\{ {\bf x}\equiv (x,y):|x|<x_0, |y|<y_0\}$ with 
specified numbers $x_0>0$ and $y>0$ (see Figure~1a). We will assume that an external temperature field
$T^e$ is created by two vertical plates $x = \pm x_0$
which are kept at different values $T_1$ and $T_2$, while the upper and lower plates $y=\pm y_0$ are thermally insulated. We assume further that
there is a material shell $(\Omega,\kappa)$ inside $D$. Here, $\Omega$ is a
circular layer $a < |{\bf x}| < b\}$, $b<\max\{x_0, y_0\}$ and $\kappa$ is the thermal conductivity tensor of the inhomogeneous anisotropic medium filling the domain $\Omega$. It is also assumed that the
interior $\Omega_i$: $|{\bf x}|<a$ and exterior $\Omega_e$: $|{\bf x}|>b$ of $\Omega$ are
filled with homogeneous media having constant thermal conductivity $k_b > 0$  (see Figure~1b).

In this case, the direct conduction problem consists of determination of a triplet of functions, namely $T_i$ in $\Omega_i$, $T$ in $\Omega$, and $T_e$ in $\Omega_e$, which satisfy the equations
\begin{equation}
\label{eq1}
k_b \, \Delta T_i = 0 \; \mbox{ in }\; \Omega_i,
\end{equation}
\begin{equation}
{\rm div}\, (\kappa \, {\rm grad} \, T) = 0\; \mbox{ in }\; \Omega,
\end{equation}
\begin{equation}
k_b \, \Delta T_e = 0 \; \mbox{ in } \; \Omega_e, 
\end{equation}
obey the following boundary conditions on the boundary $\partial D$ of $D$
\begin{equation}
{T_e}|_{x= -x_0} = T_1, \;\; {T_e}|_{x=x_0} = T_2, \;\; \frac{\partial {T_e}}{\partial y}|_{y=\pm y_0} = 0
\end{equation}
and the matching conditions on internal $\Gamma_i$ and external $\Gamma_e$ components of the boundary $\Gamma$ of the shell $\Omega$ having the form
\begin{equation}
T_i = T, \; \; k_b \frac{\partial T_i}{\partial n} = (\kappa \nabla T) \cdot {\bf n} \mbox{ on }\Gamma_i, \;\;\;
T_e = T, \; \; k_b \frac{\partial T_e}{\partial n} = (\kappa \nabla T) \cdot {\bf n} \mbox{ on } \Gamma_e.
\label{eq6}
\end{equation}
\begin{figure}[h!]
\begin{minipage}[h]{0.49\linewidth}
\center{\includegraphics[width=0.9\linewidth]{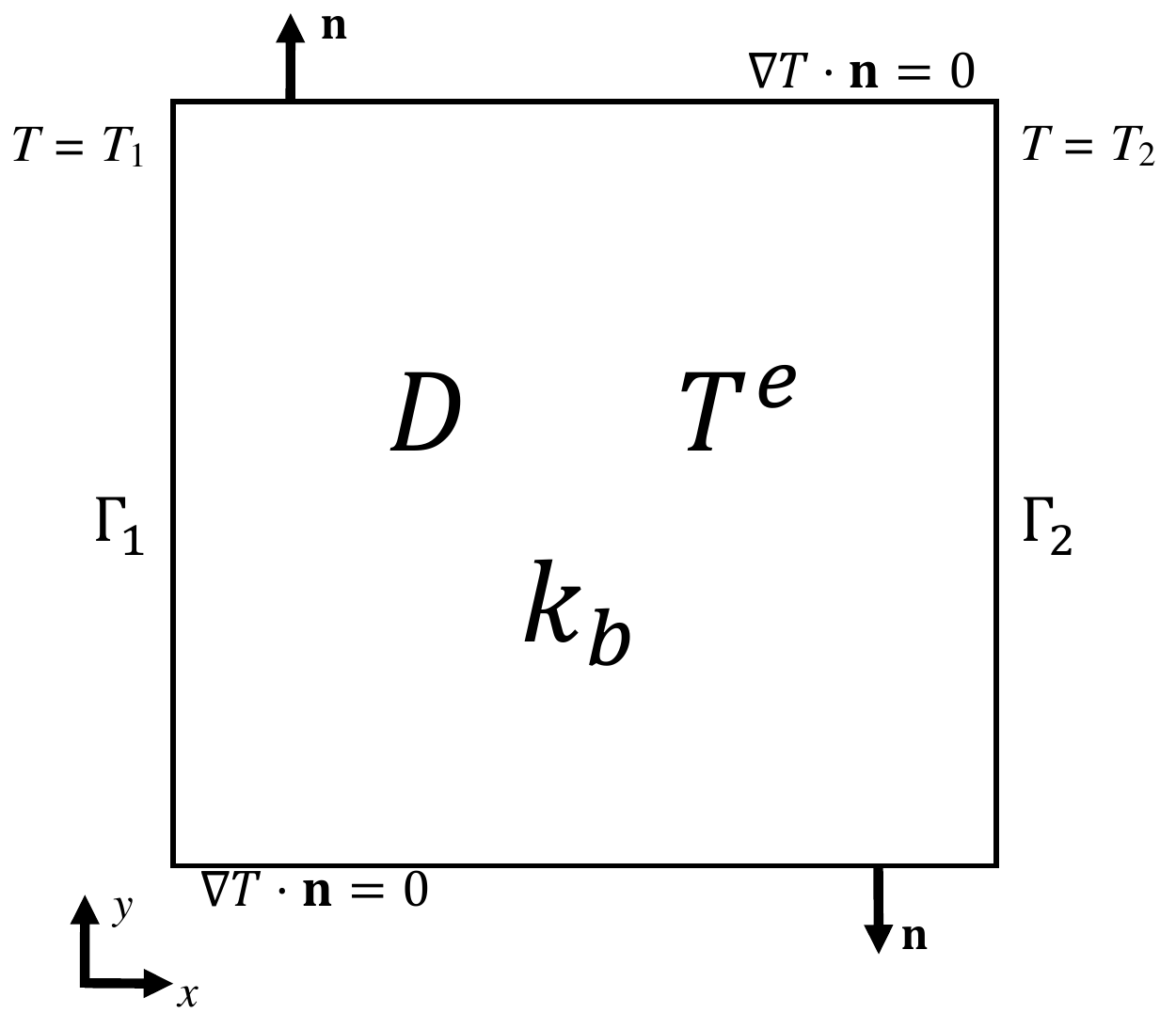}}
\end{minipage}
\hfill
\begin{minipage}[h!]{0.49\linewidth}
\center{\includegraphics[width=0.9\linewidth]{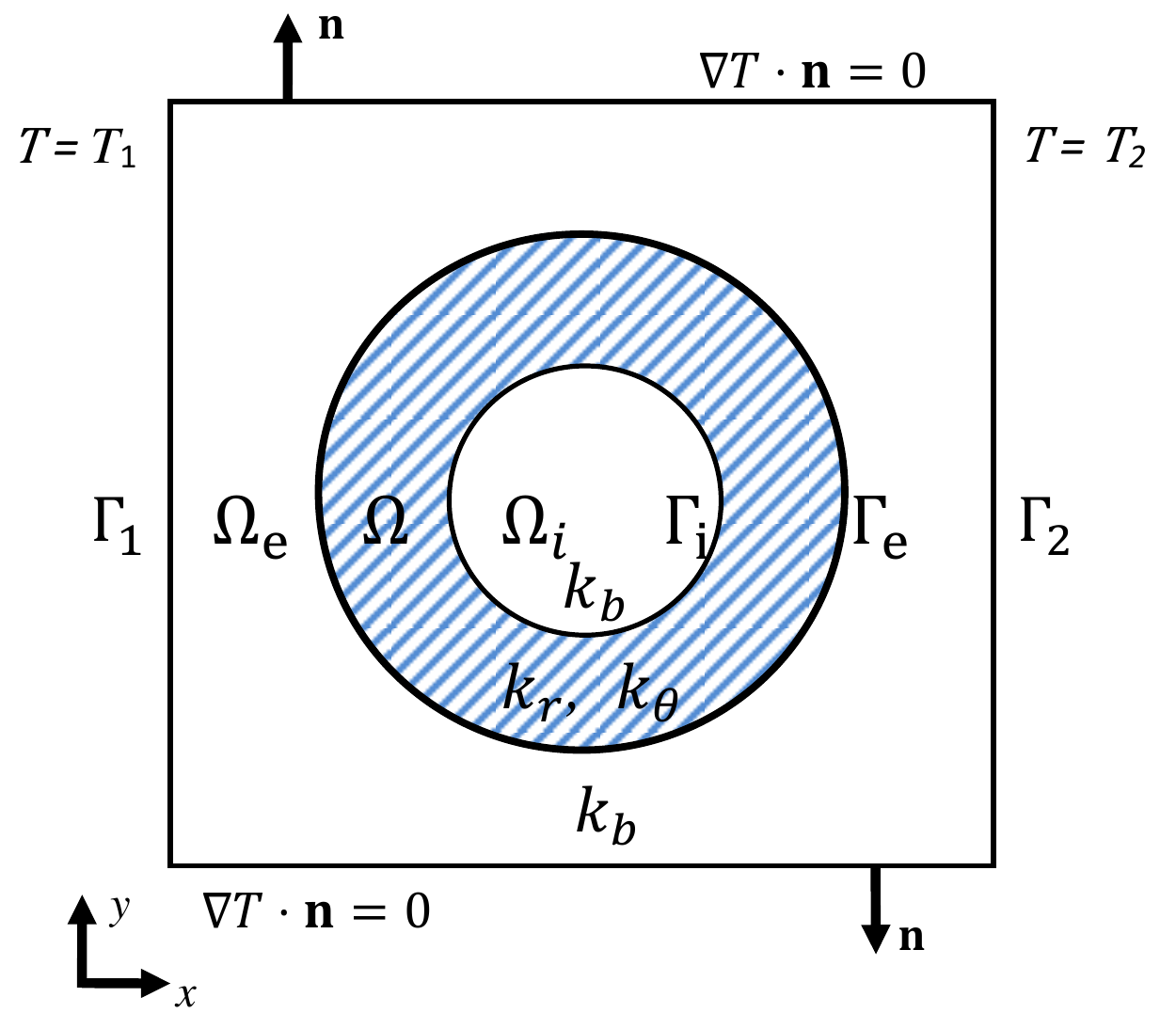}}
\end{minipage}
\begin{center}
\large a) \hspace{7cm} b)
\end{center}
\caption{Two geometries of the problem: a) without a shell;  b) with a shell.}
\end{figure}

As already mentioned, our goal is to analyze the cloaking problems for the model (\ref{eq1})--(\ref{eq6}).
Let us remind that the general inverse cloaking problem in the exact formulation
consists in finding thermal conductivity $\kappa$ providing the
following cloaking conditions \cite{Guen12,Al18}:
\begin{equation}
\label{eqCC}
T_e= T^e  \mbox{ in } \; \Omega_e, \;\; \nabla T_i= 0 \; \mbox{ in } \; \Omega_i.
\end{equation}
These conditions imply that temperature field must be equal to the external field $T^e$  in $\Omega_e$ and 
the gradient $\nabla T_i$ must be equal to zero in $\Omega_i$.
Besides this general problem we will also study the external cloaking problem. It consists in finding 
conductivity $\kappa$ providing the fulfillment of the first condition in (\ref{eqCC}).

Preliminarily we will study some properties of solutions to direct problem (\ref{eq1})--(\ref{eq6}).
We will assume firstly that the tensor $\kappa$ is diagonal in polar coordinates $r$, $\theta$ and its diagonal components (radial and polar conductivities) $k_r$ and $k_\theta$ are the positive constants. 
This condition is widely used when analysing two-dimensional problems of designing annular functional devices. 
Moreover, in the particular case when the boundary functions $T_1$ and $T_2$ are constants, $k_b={\rm const}$ and 
the so called admissibility condition $k_r k_\theta=k_b^2$ is satisfied, problem (\ref{eq1})--(\ref{eq6}) has an exact 
solution which can be found by Fourier method. This solution $(T_i,T,T_e)$ is described by 
\[
T_i (r, \theta) = \frac{T_0}{x_0} {\cal K}(s) \, r\cos \theta 
+\frac{T_1 +T_2}{2} \; {\rm in} \; \Omega_i,  \;
T (r, \theta) =  \frac{T_0}{x_0} \left( \frac{r}{b} \right)^{s-1} r\cos \theta 
+\frac{T_1+T_2}{2} \; {\rm in} \; \Omega,
\]
\begin{equation}
\label{eq7}
T_e (r, \theta) =  \frac{T_0}{x_0} r \cos  \theta +\frac{T_1+T_2}{2} \; \mbox{ in } \; \Omega_e
\end{equation}
where
\begin{equation}
\label{eq7a}
s=\sqrt{\frac{k_\theta}{k_r}}, \;\;\; {\cal K}(s) =\left(\frac{a}{b}\right)^{s-1}, 
\; T_0=\frac{T_2-T_1}{2}.
\end{equation}
The parameter $s=\sqrt{k_\theta/k_r}$ characterizes the degree of anisotropy of the respective homogeneous shell 
$(\Omega, k_r, k_\theta)$. In the particular case when $s=1$ (provided that $k_r=k_\theta = k_b$ so that the entire medium filling the domain $D$ is homogeneous and isotropic as shown in Figure~1a), formula (\ref{eq7}) is transformed into the unified formula
\begin{equation}
T^e (x)=  \frac{T_0}{x_0} r \cos  \theta +\frac{T_1+T_2}{2} = T_0 \frac{x}{x_0} +\frac{T_1+T_2}{2},
\label{eq8}
\end{equation}
which describes the external applied field $T^e (x)$. Just this field $T^e$ is used to detect objects located in domain $D$. 

Assuming that $s\geq 1$ let us define function ${\cal M}(s)=1-{\cal K}(s)$ where ${\cal K}(s)$ is given in (\ref{eq7a}). Since $a<b$ we derive from (\ref{eq7a}) that
\begin{equation}
\label{new8a}
0\leq {\cal K} (s)\leq 1, \;\; 
0\leq {\cal M}(s)\leq 1, \; \; {\cal M}(s) + {\cal K}(s) = 1 \mbox{ for } s \geq 1.
\end{equation}
Moreover, we have
\[
{\cal K} (s)\to 1, \; {\cal M} (s)\to 0 \; {\rm as} \; s\to 1, \; 
{\cal K} (s)\to 0, \; {\cal M} (s)\to 1 \; {\rm as} \; s\to \infty.
\]
Also, it follows from (\ref{eq7}), (\ref{eq8}) that external and internal fields $T_e$ and $T_i$ corresponding to the homogeneous shell $(\Omega,k_r,k_\theta)$ satisfy conditions
\begin{equation}
\label{new8b}
T_e= T^e, \; |\nabla T_e|= |\nabla T^e|= \frac{|T_0|}{x_0} \; \mbox{ in } \; \Omega_e, 
|\nabla T_i|= \frac{|T_0|}{x_0} {\cal K} (s) \mbox{ in } \Omega_i.
\end{equation}

Using the above mentioned properties of functions ${\cal M}(s)$ and ${\cal K}(s)$ and taking into account terminology of \cite[Ch.~4]{Mono16} we will refer below to 
${\cal M}(s)$ and ${\cal K}(s)$ as the cloaking efficiency and visibility measure of the respective cloak $(\Omega,k_r,k_\theta)$. 
We emphasize that cloaking efficiency of the homogeneous anisotropic shell $(\Omega,k_r,k_\theta)$ increases while the visibility measure decreases with an increase 
of the anisotropy parameter $s=\sqrt{k_\theta/ k_r}$.

In the limit as $s \to \infty$ we obtain an exact cloaking shell with maximum cloaking efficiency ${\cal M} = 1$ and minimum visibility measure ${\cal K}=0$. 
On contrary, in the limit as $s\to 1$, we have  a shell $(\Omega, k_r, k_\theta)$ with minimal cloaking efficiency ${\cal M}(1)= 0$ but maximum visibility measure ${\cal K}(1)=1$. 

\section{Control problems for layered shell}

Now we consider more complicated scenario when $(\Omega, k_r, k_\theta)$ is a layered shell consisting of $M$ concentric circular layers
$\Omega_j =\{ r_{j-1}<r=|{\bf x}|<r_j\}$, $j = 1,2,\ldots,M$, where $r_0 = a$, $r_M = b$. Each of these layers is filled with a homogeneous and (generally) anisotropic medium, described by constant conductivities 
$k_{rj} > 0$ and $k_{\theta j} > 0$, $j = 1,2,\ldots,M$. 
The parameters $k_r$ and $k_\theta$ of this layered shell are given by
\begin{equation}
\label{eq10}
k_r ({\bf x}) = \sum_{j=1}^M k_{rj} \chi_j ({\bf x}), \; \;
k_\theta ({\bf x}) = \sum_{j=1}^M k_{\theta j} \chi_j ({\bf x}).
\end{equation}
Here,  $\chi_j ({\bf x})$ is a characteristic function of layer $\Omega_j$,
which is equal to $1$ inside of $\Omega_j$ and $0$ outside of $\Omega_j$, 
while $k_{rj}$ and $k_{\theta j}$ are some constant unknown coefficients. It should be noted that the case when $k_{rj}=k_{\theta j}$ corresponds to $M$-layered isotropic cloak.

We remind that our purpose is the numerical analysis of inverse problems for 2D model of heat transfer (\ref{eq1})--(\ref{eq6}) 
arising when developing technologies of designing thermal cloaking devices. In the case of a layered shell with parameters (\ref{eq10}) these problems consist of finding unknown coefficients $k_{rj}$ and $k_{\theta j}$, $j=1,2,...,M$ in 
(\ref{eq10}) forming a $2M$-dimensional vector ${\bf k}=(k_{r1},k_{\theta 1},...,k_{rM},k_{\theta M})$ from the cloaking conditions. 
We will refer to ${\bf k}$ as conductivity vector because it consists of conductivities $k_{rj}$, $k_{\theta j}$
of separate sublayers $\Omega_j$, $j=1,2,...,M$ comprising the $M$-layered shell $(\Omega,{\bf k})$.
In the particular case of the $M$-layered isotropic shell when  $k_{rj}=k_{\theta j}=k_j>0$
conductivities of separate sublayers form a $M$-dimensional vector 
${\bf k}=(k_1,,...,k_M)$.
 
By optimization method our inverse problems are reduced to extremum problems of minimization of certain cost functionals which adequately correspond to inverse problems of designing devices for approximate cloaking \cite{AlLeTe17s}. 
 In order to formulate the control problems under study, we denote by 
$T [{\bf k}] \equiv T [k_{r 1}, k_{\theta 1}, \ldots , k_{r M}, k_{\theta M}]$ the solution to the direct problem (\ref{eq1})--(\ref{eq6}) corresponding to parameters 
$(k_{r j}, k_{\theta j})$ in $\Omega_j$, $j=1,2,\ldots,M$, and to conductivity $k_b$ in $\Omega_i$ and $\Omega_e$. 
We will assume below that the vector ${\bf k} = (k_{r 1}, k_{\theta 1}, \ldots , k_{r M}, k_{\theta M})$ belongs to the bounded set
\begin{equation}
\label{eqS}
K= \{ {\bf k} \equiv (k_{r 1}, k_{\theta 1}, \ldots , k_{r M}, k_{\theta M}): \, 
m_r  \le k_{rj} \le M_r, \; m_\theta  \le  k_{\theta j}  \le M_\theta, \, j=1,\ldots,M \}
\end{equation}
to which we will refer to as a control set.  Here given positive constants  $m_r$, $m_\theta$ and $M_r$,  $M_\theta$ 
are lower and upper bounds of the control set $K$. Let us define two cost functionals
\begin{equation}
\label{eq11}
J_e ({\bf k}) = \frac{\|T[{\bf k}]-T^e\|_{L^2(\Omega_e)}}{\|T^e\|_{L^2(\Omega_e)}}, 
 \; J_i ({\bf k}) = \frac{\| \nabla T [{\bf k}]\|_{L^2(\Omega_i)}}{\| \nabla T^e\|_{L^2(\Omega_i)}},
\end{equation} 
where, in particular, 
\[
\|T^e\|^2_{L^2(\Omega_e)} = \int_{\Omega_e} |T^e|^2 d {\bf x}, \;\;
\|T[{\bf k}]-T^e\|^2_{L^2(\Omega_e)} = \int_{\Omega_e} |T[{\bf s}]-T^e|^2 d {\bf x}, 
\]
and formulate following two control problems:
\begin{equation}
\label{eqJe}
J_e ({\bf k})   \to \inf, \;\;  {\bf k}   \in K,
\end{equation} 
\begin{equation}
\label{eqJ}
J({\bf k})= \frac{J_e ({\bf k})+J_i ({\bf k})}{2}    \to \inf, \;\; {\bf k}   \in K. 
\end{equation} 
It follows from previous analysis (see also \cite{Mono16}) that problem (\ref{eqJe}) is aimed to finding an approximate (optimal) solution of the external cloaking problem while problem  (\ref{eqJ}) is aimed to finding an approximate solution of the general cloaking problem.

Two particular cases of problems (\ref{eqJe}) and (\ref{eqJ}) are of particular interest.
The first case is a single-layer shell with constant parameters $k_r^*$, $k_\theta^*$
satisfying the admissibility condition $k_r^* k_\theta^* =k_b^2$. As was mentioned before,  in this case there exists 
the exact solution $(T_i,T,T_e)$ of problem  (\ref{eq1})--(\ref{eq6}) which is described by 
(\ref{eq7}), (\ref{eq7a}) and satisfies the conditions (\ref{new8b}). Using (\ref{new8b})
one easily shows that for any admissible pair $(k_r^*,k_\theta^*)$ the following relations hold:
\begin{equation}
\label{eqK}
J_e(k_r^*,k_\theta^*)=0,  \;\; J_i(k_r^*,k_\theta^*)= {\cal K}(s^*), \;\;
J(k_r^*,k_\theta^*)= 0.5 \, {\cal K}(s^*), \;\; s^*=\sqrt{k_\theta^* / k_r^*}.
\end{equation}
Here ${\cal K}(s^*)$ is the visibility measure  of the cloak $(\Omega,k_r^*,k_\theta^*)$
connected with the cloaking efficiency by the relation
${\cal K}(s^*)=1-{\cal M}(s^*)$. From here and from (\ref{eqK}) follows that value $J_i(k_r^*,k_\theta^*)$
of the functional $J_i$ as well as value $J(k_r^*,k_\theta^*)$ of the functional $J$ 
characterizes the cloaking  efficiency of the corresponding shell $(\Omega,k_r^*,k_\theta^*)$.
The smaller value $J(k_r^*,k_\theta^*)$ corresponds to smaller value of visibility measure 
which in turn corresponds to higher cloaking efficiency of the 
shell $(\Omega,k_r^*,k_\theta^*)$ and vice versa. 
We emphasize that the value $J({\bf k})$ of the functional $J$ on any element 
${\bf k} \equiv (k_{r 1}, k_{\theta 1}, \ldots , k_{r M}, k_{\theta M})$
can be considered as a visibility measure of the respective cloak $(\Omega,{\bf k}^{opt})$
which characterize a cloaking efficiency and in the general case for 
$M$-layered anisotropic (or isotropic) shell $(\Omega,{\bf k})$.
Therefore our goal will consist of finding a conductivity vector (an optimal solution of (\ref{eqJ}))
${\bf k}^{opt} \in K$ for which the functional $J$ takes a minimum value $J^{opt}=J({\bf k}^{opt})$
on the set $K$ and therefore the cloak 
$(\Omega,{\bf k}^{opt})$ possess a maximum cloaking efficiency.

To solve problems (\ref{eqJe}), (\ref{eqJ}) we use an algorithm based on the particle swarm optimization \cite{Po07}.
Within this method, the desired parameters determining the value of minimized functional $J$ are presented
in the form of the coordinates of the radius-vector 
${\bf k} \equiv (k_{r 1}, k_{\theta 1}, \ldots , k_{r M}, k_{\theta M})$
of some abstract particle. A particle swarm is considered to be any finite set of particles 
${\bf k}_1, \ldots,{\bf k}_N$. Within the particle swarm optimization, one sets the initial swarm 
position ${\bf k}_0^j$, $j=1,2,\ldots,N$, and the iterative displacement procedure 
${\bf k}_j^{i+1}={\bf k}_j^i+{\bf v}_j^{i+1}$ for all particles ${\bf k}_j$, which is described by the formula (see \cite{AlLeTe17c})
\begin{equation}
\label{eqV}
{\bf k}_j^{i+1}=w {\bf v}_j^i + c_1 d_1 ({\bf p}_j^i-{\bf k}_j^i)
 + c_2 d_2 ({\bf p}_g-{\bf k}_j^i).
\end{equation}
After each displacement we calculate the value $J ({\bf k}_j^{i+1})$ of the functional $J$ for the new position 
${\bf k}_j^{i+1}$, compare it to the current minimum value and, if necessary, update the personal and
global best positions ${\bf p}_j$ and ${\bf p}_g$.
In the end of this iteration process, all particles have to come at the global minimum point.

Here, ${\bf v}_j$ is the displacement vector; $w$, $c_1$ and $c_2$ are constant parameters; and $d_1$ and $d_2$ are
random variables, which are uniformly distributed over the interval $(0,1)$. Their choice was considered in
more detail in \cite{AlLeTe17c,Po07}. The subscript $j \in \{1 , 2, \ldots , N \}$ in (\ref{eqV}) denotes the particle number, and the superscript $i \in \{0 , 1, \ldots , L \}$ indicates the iteration number.

\section{Simulation results}
In this section we discuss the results which were obtained using the PSO algorithm 
for designing anisotropic single-layer and isotropic multilayer cloaking shells.
The most time-consuming part of the above-described algorithm is the calculation of the values
$J ({\bf k}_j^i)$ of the functional $J$  for the particle position ${\bf k}_j^{i+1}$
at different $i$ and $j$ values. This procedure includes two stages. At the first stage, the solution
$T[{\bf k}_j^i]$ to direct problem (\ref{eq1})--(\ref{eq6}) is calculated. To this end, we used the FreeFEM++ software package (www.freefem.org), designed for numerical solution of two- 
and three-dimensional boundary value problems by the finite element method.  After determining $T[{\bf k}_j^i]$, at the second stage we calculate the mean squared integral norms 
entering the definition of functionals $J_e$ or $J_i$ in (\ref{eq11}) using the formulas of numerical integration.

Numerical experiments were performed for values $N=25$, $L=50$, $w=0.4$, $c_1=1$, $c_2=1.5$ of 
PSO parameters and at the following given data:
\begin{equation}
\label{new15}
x_0 = y_0 = 3 \mbox{ m}, \; a = 1  \mbox{ m}, \; b = 2 \mbox{ m},   T_1 = 100 \mbox{ C}^\circ,  \; T_2 = 0 \mbox{ C}^\circ, \; k_b= k_0 \equiv 1 \mbox{ W/(m}\cdot \mbox{K)}.
\end{equation} 
The role of the external field was played by the field $T^e$ determined in (\ref{eq8}) at $T_1 = 100$~C$^\circ$,
$T_2 = 0$~C$^\circ$. It is characterized by a constant gradient $\nabla T^e$ with a modulus $|\nabla T^e| = |T_0| / x_0$ and with straight isolines $T^e={\rm const}$, oriented perpendicular to the $x$ axis (see Figure~2).

\begin{figure}[b!]
\begin{center}
\includegraphics[width=0.61\linewidth, height=0.5\linewidth]{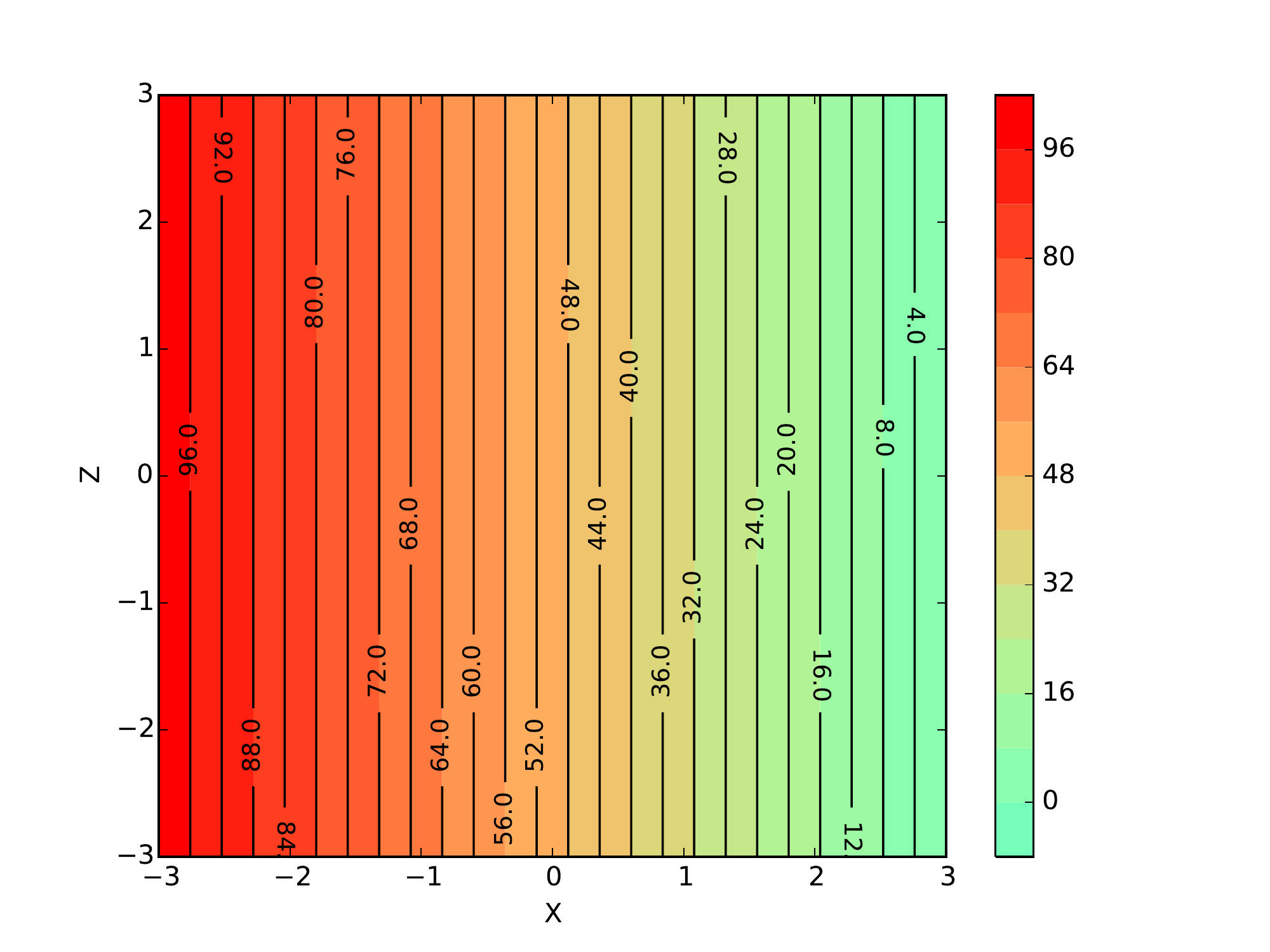} 
\caption{Isolines for the external field $T^e$.}
\end{center}
\end{figure}

Firstly, we solve the design problem for a single-layer anisotropic shell 
$(\Omega,k_r,k_\theta)$ assuming that the background material in $\Omega_i$ and $\Omega_e$
is glass ($k_b=k_0$) while lower and upper bounds in (\ref{eqS}) are determined by
\begin{equation}
\label{new16}
m_r=0.05 \, k_0, \; M_r=1.0 \, k_0, \;\; m_\theta=5  \, k_0,  \; M_\theta=15  \, k_0. 
\end{equation}
Numerical solution of control problem (\ref{eqS}), (\ref{eqJe}) 
with the help of PSO algorithm gives after 50 iterations the following results:
\[
k_r^{opt}=0.18  \, k_0, \;\; k_\theta^{opt}=5.54  \, k_0, \;\; 
J_e^{opt} =2.27 \times 10^{-6}, \; J_i^{opt} =4.31 \times 10^{-2}, \;
J^{opt} =2.15 \times 10^{-2}.
\]
Isolines for corresponding temperature field $T^{opt} = T[k_r^{opt}, k_\theta^{opt}]$ are shown in Figure~3a.
The isolines are straight as if there is no shell in domain $D$, and it creates the illusion of the absence of the shell $\Omega$ for an external observer. 

As the second test we solve (under the mentioned conditions (\ref{new15}), (\ref{new16})  to the data) the problem (\ref{eqS}), (\ref{eqJ}) using the developed PSO algorithm. After 50 iterations we obtain the following results:
\[
k_r^{opt}=7.42 \times 10^{-2}   \, k_0, \; k_\theta^{opt}=13.46   \, k_0, \;\;
J_e^{opt} =7.92 \times 10^{-5}, \; J_i^{opt} =1.75 \times 10^{-4}, \;
J^{opt} =1.27 \times 10^{-4}.
\]

\begin{figure}[h!]
\begin{minipage}[h]{0.50\linewidth}
\center{\includegraphics[width=0.95\linewidth]{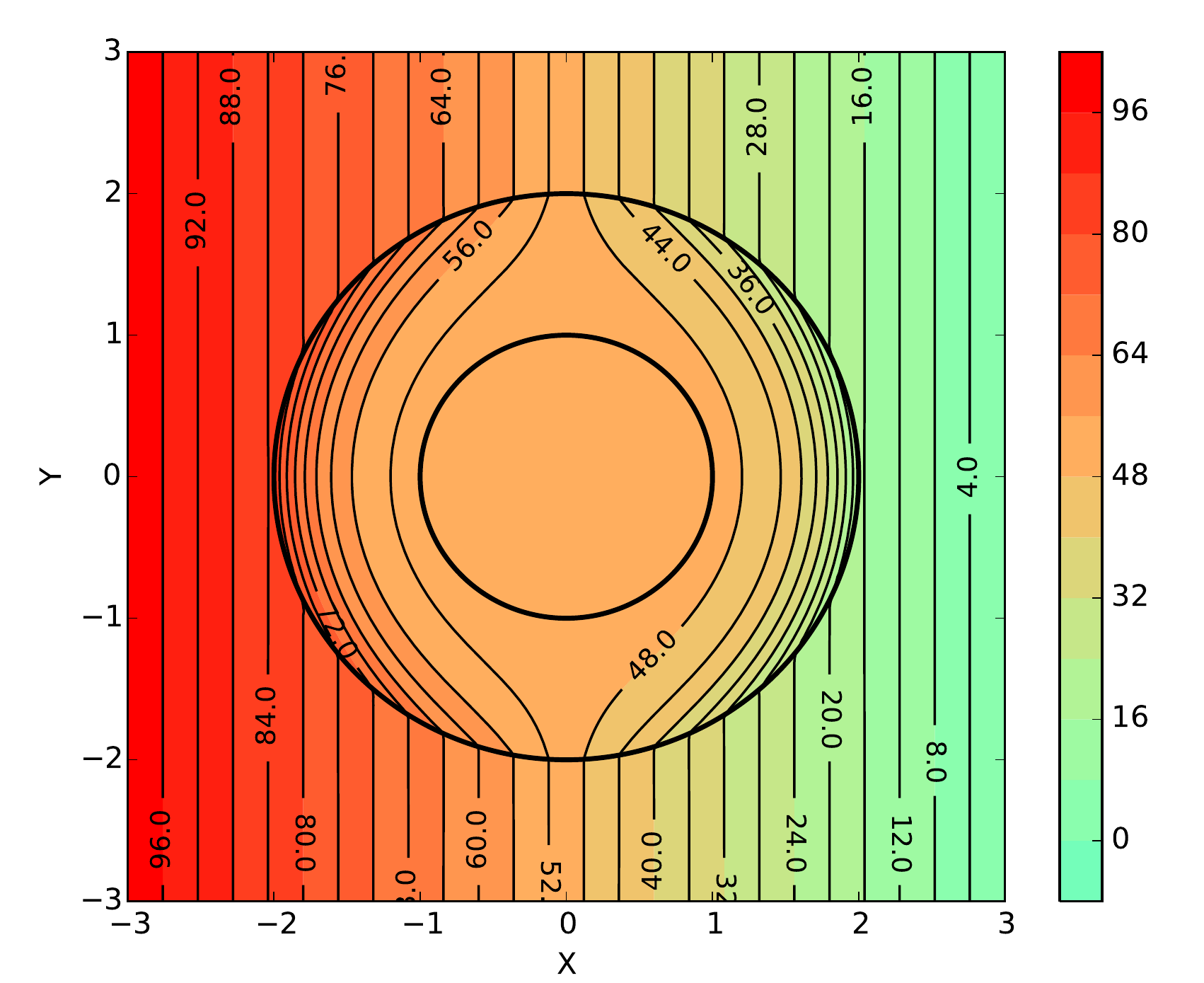}}
\end{minipage}
\hfill
\begin{minipage}[h!]{0.50\linewidth}
\center{\includegraphics[width=0.97\linewidth]{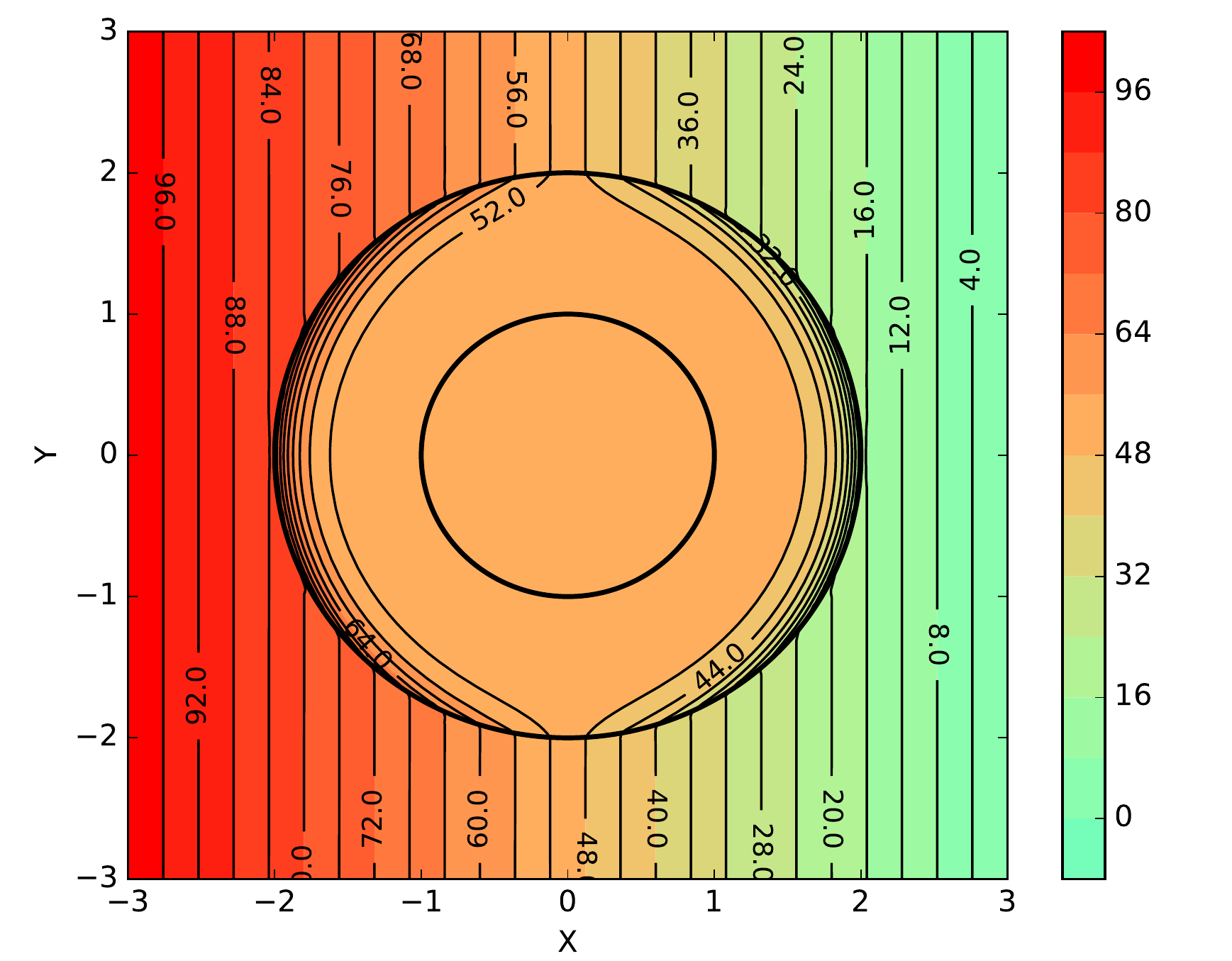}}
\end{minipage}
\begin{center}
\large a) \hspace{8cm} b)
\end{center}
\caption{Isolines of temperature field for  anisotropic cloaking shell: a) $J_e$; b)  $J$.}
\end{figure}

\noindent The isolines of calculated temperature field $T^{opt} = T[k_r^{opt}, k_\theta^{opt}]$ are plotted in Figure~3b.
Again we note that the isolines are close to isolines in $\Omega_e$ in Figure~2 and Figure~3a.
It should be noted however that this high cloaking efficiency is achieved in both cases due to the high anisotropy
of respective cloak $(\Omega,k_r^{opt},k_\theta^{opt})$. It is characterized by anisotropy coefficient 
$s^{opt}=\sqrt{k_\theta^{opt}/ k_r^{opt}}$ which is equal to $5.5$ for the solution 
of problem (\ref{eqS}), (\ref{eqJe}) and  is equal to $13.5$ for the solution of problem (\ref{eqS}), (\ref{eqJ}).

As mentioned above the technical realization of highly anisotropic cloaking shells is associated with great difficulties because appropriate materials are not found in nature and should be created using metamaterials approach.
One of the ways of overcoming these difficulties consist of using 
layered shell consisting of $M$ concentric circular layers
$\Omega_j =\{ r_{j-1}<r=|{\bf x}|<r_j\}$, $j = 1,2,\ldots,M$, where $r_0 = a$, $r_M = b$. Each of these layers is filled with a homogeneous and isotropic medium, described by constant conductivity 
$k_j$, $j = 1,2,\ldots,M$.
Set ${\bf k}=(k_1, k_2, \ldots, k_M)$ and define the control set $K$ by 
\begin{equation}
\label{eqS2}
K=\{ {\bf k}=(k_1, k_2, \ldots, k_M): \,  k_{min} \le k_j \le  k_{max}, \; j=1,\ldots,M \}. 
\end{equation}

We begin with analysis of cloaking efficiency of a multilayer cloaking shell proposed in \cite{Han13}. 
It is composed  of alternate layers of two materials (wood and stainless steel), whose conductivities are defined by the following parameters: $k_1=k_3=\ldots=k_{M-1}=5 \times 10^{-2} \, k_0$, 
$k_2=k_4=\ldots=k_M=20 \, k_0$ and, besides, $k_b=k_0$. 
Since all cloak parameters are specified, it remains to us  to determine the cloaking efficiency 
of the corresponding cloaking shell $(\Omega, {\bf k})$ by calculating the values $J({\bf k})$, 
$J_i({\bf k})$ and $J_e({\bf k})$.
These values together with $k_M = 20 \, k_0$ are presented for six different shells ($M=2$, $4$, $6$, $8$, $10$ and $12$)  in Table~1. 
The isolines of corresponding temperature field for a six-layer shell ($M=6$) are shown in Figure~4a.
It is seen that $J({\bf k})$ takes the value $1.6 \times 10^{-2}$ for $M=12$. 
Besides, the isolines outside cloak are curved. 
These results indicate low cloaking efficiency of the above designed cloak $(\Omega,{\bf k})$.

\begin{table}[th]
\caption{Numerical results for unoptimized multilayer shells 
($k_b = k_0$, $k_{min} = 0.05  \, k_0$,  $k_{max} = 20 \, k_0$).}
\label{table1}
\begin{center}
\begin{tabular}{ccccc}
\hline
$M$   & $k_M / k_0$ & $J({\bf k})$   & $J_i({\bf k})$   & $J_e({\bf k})$  \\
\hline
  2 \rule{0cm}{4mm}    & $20$  & $1.1 \times 10^{-1}$ & $8.8 \times 10^{-2}$ & $1.2 \times 10^{-1}$ \\
  4    & $20$  & $4.7 \times 10^{-2}$ & $1.6 \times 10^{-2}$ & $7.8 \times 10^{-2}$ \\
  6    & $20$  & $3.2 \times 10^{-2}$ & $8.3 \times 10^{-3}$ & $5.6 \times 10^{-2}$ \\
  8    & $20$  & $2.4 \times 10^{-2}$ & $5.7 \times 10^{-3}$ & $4.2 \times 10^{-2}$ \\
  10   & $20$  & $1.9 \times 10^{-2}$ & $4.4 \times 10^{-3}$ & $3.4 \times 10^{-2}$ \\
  12   & $20$  & $1.6 \times 10^{-2}$ & $4.1 \times 10^{-3}$ & $2.8 \times 10^{-2}$ \\
\hline
\end{tabular}
\end{center}
\end{table}

\begin{figure}[h!]
\begin{minipage}[h]{0.50\linewidth}
\center{\includegraphics[width=1.1\linewidth,height=0.9\linewidth]{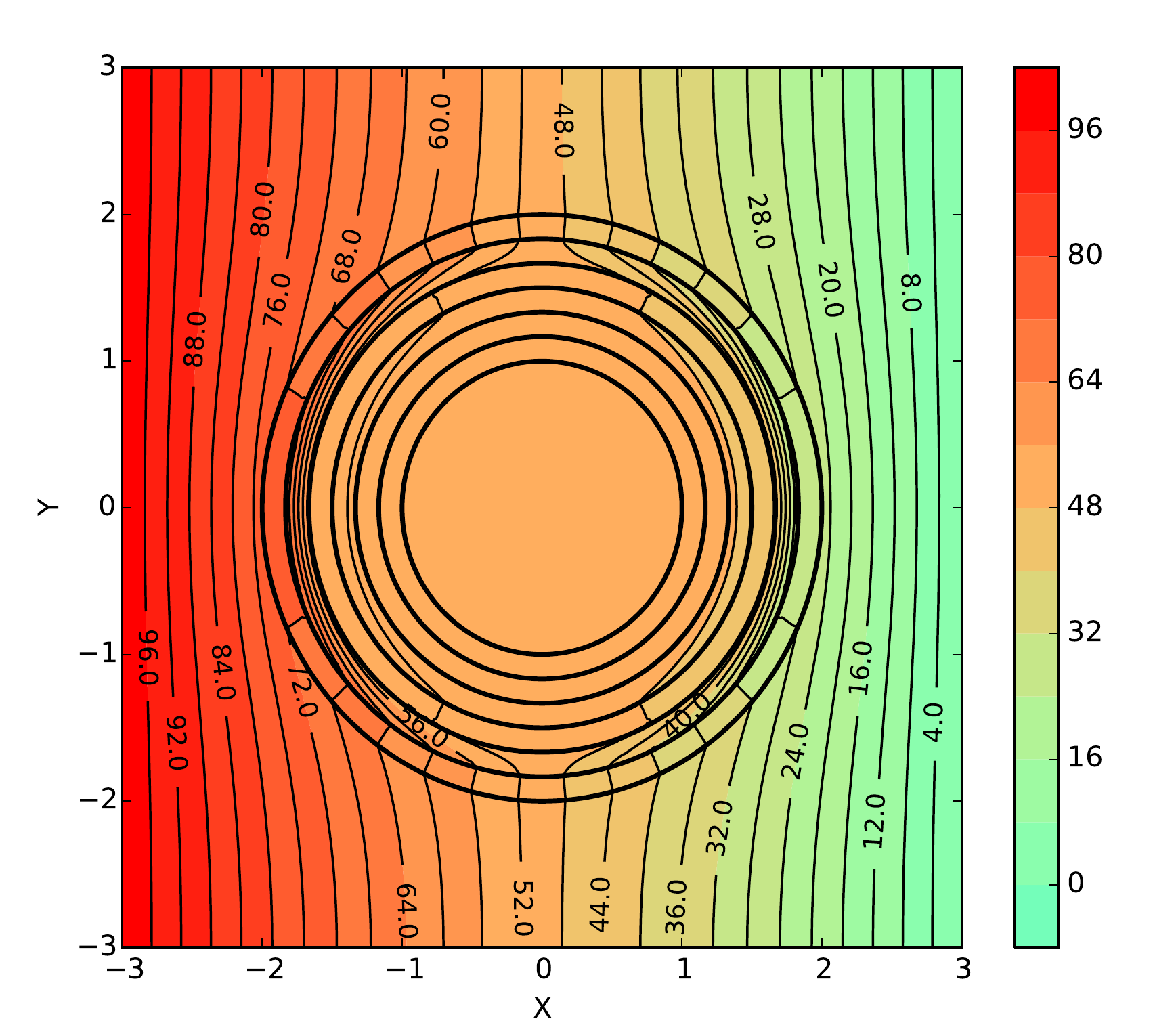}}
\end{minipage}
\hfill
\begin{minipage}[h!]{0.50\linewidth}
\center{\includegraphics[width=1.05\linewidth,height=0.87\linewidth]{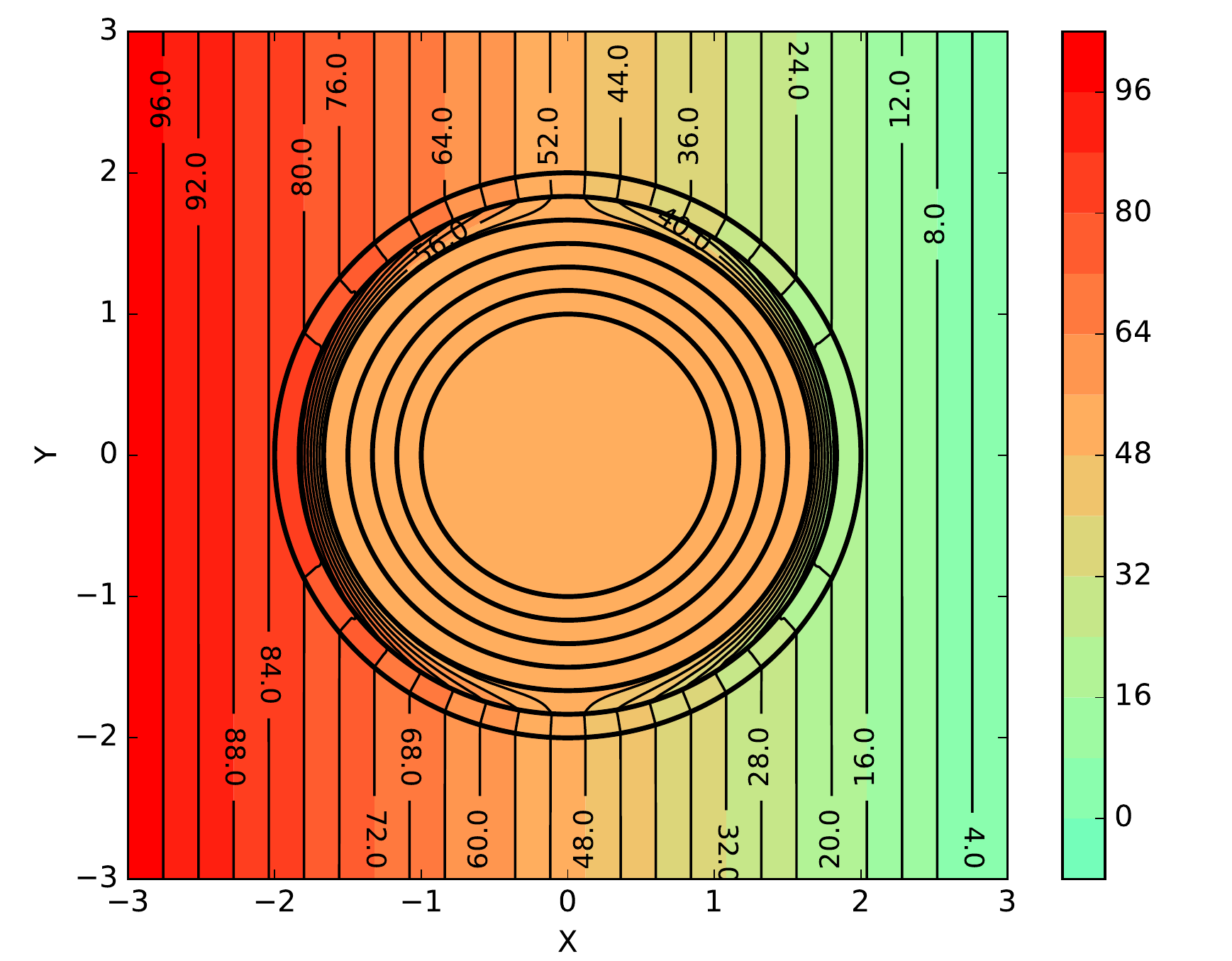}}
\end{minipage}
\begin{center}
\large a) \hspace{8cm} b)
\end{center}
\caption{Temperature isolines for six-layer shell: a)  unoptimized; b)  optimized.}
\end{figure}

In order to design a cloaking shell with higher cloaking efficiency 
we apply the optimization method.
To this end we solve the control problem  (\ref{eqJ}), (\ref{eqS2}) for the case when
$k_{min}=5 \times 10^{-2} \, k_0$, $k_{max}=20 \, k_0$ and $k_b=k_0$. 
Our optimization analysis  using PSO algorithm showed that optimal values $k_i^{opt}$ of  all parameters $k_i$ 
with odd indices $i= 3,5,\ldots,M-1$ coincide with  $k_{min}$,  optimal values of parameters 
$k_2^{opt},\ldots,k_{M-2}^{opt}$ with even indices coincide with $k_{max}$
while $k_M^{opt}$ changes from $20 \, k_0$ at $M=2$ to  $8.15 \, k_0$ at $M=12$.
Thus we have 
\begin{equation}
\label{eqN}
k_1^{opt}=k_3^{opt}=\ldots=k_{M-1}^{opt}= k_{min}, \;\;
k_2^{opt}=k_4^{opt}=\ldots=k_{M-2}^{opt}=k_{max}
\end{equation}
while optimal value $k_M^{opt}$ of the last control $k_M$ and corresponding values $J^{opt}$, $J_i^{opt}$ and $J_e^{opt}$ for different $M=2$, $4$,$...$, $12$ are presented in Table~2. 
Isolines of corresponding field $T^{opt}$ for a six-layer shell ($M=6$) are plotted in Figure~4b.

\begin{table}[th]
\caption{Numerical results for optimized multilayer shells 
($k_b = k_0$, $k_{min} = 0.05 \, k_0$,  $k_{max} = 20 \, k_0$).}
\label{table2}
\begin{center}
\begin{tabular}{ccccc}
\hline
$M$   & $k_M^{opt} / k_0$ & $J^{opt}$   & $J_i^{opt}$   & $J_e^{opt}$  \\
\hline
  2 \rule{0cm}{4mm}    & $20$  & $1.1 \times 10^{-1}$ & $8.8 \times 10^{-2}$ & $1.2 \times 10^{-1}$ \\
  4    & $5.32$  & $1.4 \times 10^{-2}$ & $2.7 \times 10^{-2}$ & $8.7 \times 10^{-5}$ \\
  6    & $6.57$  & $5.5 \times 10^{-3}$ & $1.1 \times 10^{-2}$ & $5.1 \times 10^{-5}$ \\
  8    & $6.72$  & $3.8 \times 10^{-3}$ & $7.6 \times 10^{-3}$ & $4.7 \times 10^{-5}$ \\
  10   & $7.76$  & $2.8 \times 10^{-3}$ & $5.6 \times 10^{-3}$ & $2.3 \times 10^{-5}$ \\
  12   & $8.04$  & $2.4 \times 10^{-3}$ & $4.9 \times 10^{-3}$ & $7.8 \times 10^{-6}$ \\
\hline
\end{tabular}
\end{center}
\end{table}

Analysis of Table 2 shows that the optimal values $J^{opt}$, $J_i^{opt}$ and $J_e^{opt}$ of all cost functionals decrease with increasing the number of layers $M$. 
In particular, $J^{opt}$ decreases from $1.1 \times 10^{-1}$ at $M=2$ to $2.4 \times 10^{-3}$ at $M=12$.
Temperature isolines outside the cloak in Figure~4a are straight as if there is no shell in the domain.
This indicates the relatively high cloaking efficiency of the shell $(\Omega,{\bf k}^{opt})$.

The high cloaking efficiency of the designed shell can be explained, in fact, by optimal choice of the last layer material.
It should be noted that the quality of the optimal solution, which is characterized by the optimal value $J^{opt}$ of the cost functional $J$, depends very substantially on the control set $K$ defined in (\ref{eqS2}).
If, in particular, one extends the set $K$, 
then this inevitably results in a decrease in the value $J^{opt}$ and consequently to increase in the efficiency of the respective cloaking shell.
To see this, we consider as second test the scenario when again $k_{min} = 0.05 \, k_0$, 
$k_b = k_0$ while $k_{max} = 236 \, k_0$ (corresponds to aluminium). 
Applying PSO algorithm again results in relations (\ref{eqN}) at $k_{min} = 0.05 \, k_0$ and $k_{max} = 236 \, k_0$
for all $k_i^{opt}$ except $k_M^{opt}$ and to value $k_M^{opt}$ which takes different values depending on $M$.
Optimal value $k_M^{opt}$ of the last control and corresponding values $J^{opt}$, $J_i^{opt}$ and $J_e^{opt}$ for different $M=2$, $4$,$...$, $12$ are presented in Table~3. 
We note, in particular, that $k_{10}^{opt}=2.85 \, k_0$,  $J^{opt}=1.9 \times 10^{-6}$ for $M=10$ and 
$k_{12}^{opt}=0.75$, $J^{opt}=1.1 \times 10^{-6}$ for $M=12$.
More detailed results of the functional $J$ iterative minimization for the second test are presented in Figure~5  that shows dependence of the current value $J_{min} (i)$ on the iteration number $i$ for different numbers of shell layers $M = 4, 6, 8, 10, 12$. Here $J_{min} (i)$ is the minimum value of the functional $J$ at the $i$-th iteration.
In all considered cases the algorithm converges faster then in 20 iterations. 
It is clearly seen that values of functional $J$  in Table~3 are substantially less than in Table~2.

\begin{table}[h!]
\caption{\label{table23} Numerical results for optimized multilayer shells 
($k_b = k_0$, $k_{min} = 0.05 \, k_0$, $k_{max} = 236 \, k_0$).}
\begin{center}
\begin{tabular}{ccccc}
\hline
$M$ &  $k_M^{opt} / k_0$  & $J^{opt}$   & $J_i^{opt}$   & $J_e^{opt}$  \\
\hline
  2 \rule{0cm}{4mm}  & $236$  & $1.5 \times 10^{-1}$ & $1.0 \times 10^{-2}$ & $2.0 \times 10^{-1}$ \\
  4  & $5.15$  & $1.3 \times 10^{-3}$ & $2.6 \times 10^{-3}$ & $4.2 \times 10^{-6}$ \\
  6  & $5.67$  & $1.1 \times 10^{-4}$ & $1.3 \times 10^{-4}$ & $9.3 \times 10^{-5}$ \\
  8  & $4.55$  & $2.4 \times 10^{-5}$ & $1.7 \times 10^{-5}$ & $3.1 \times 10^{-5}$ \\
  10 & $2.85$  & $1.9 \times 10^{-6}$ & $2.4 \times 10^{-6}$ & $1.4 \times 10^{-6}$ \\
  12 & $0.75$  & $1.1 \times 10^{-6}$ & $7.5 \times 10^{-7}$ & $1.5 \times 10^{-6}$ \\
\hline
\end{tabular}
\end{center}
\end{table}

\begin{figure}[th!]
\begin{center}
\includegraphics[width=0.75\linewidth, height=0.5\linewidth]{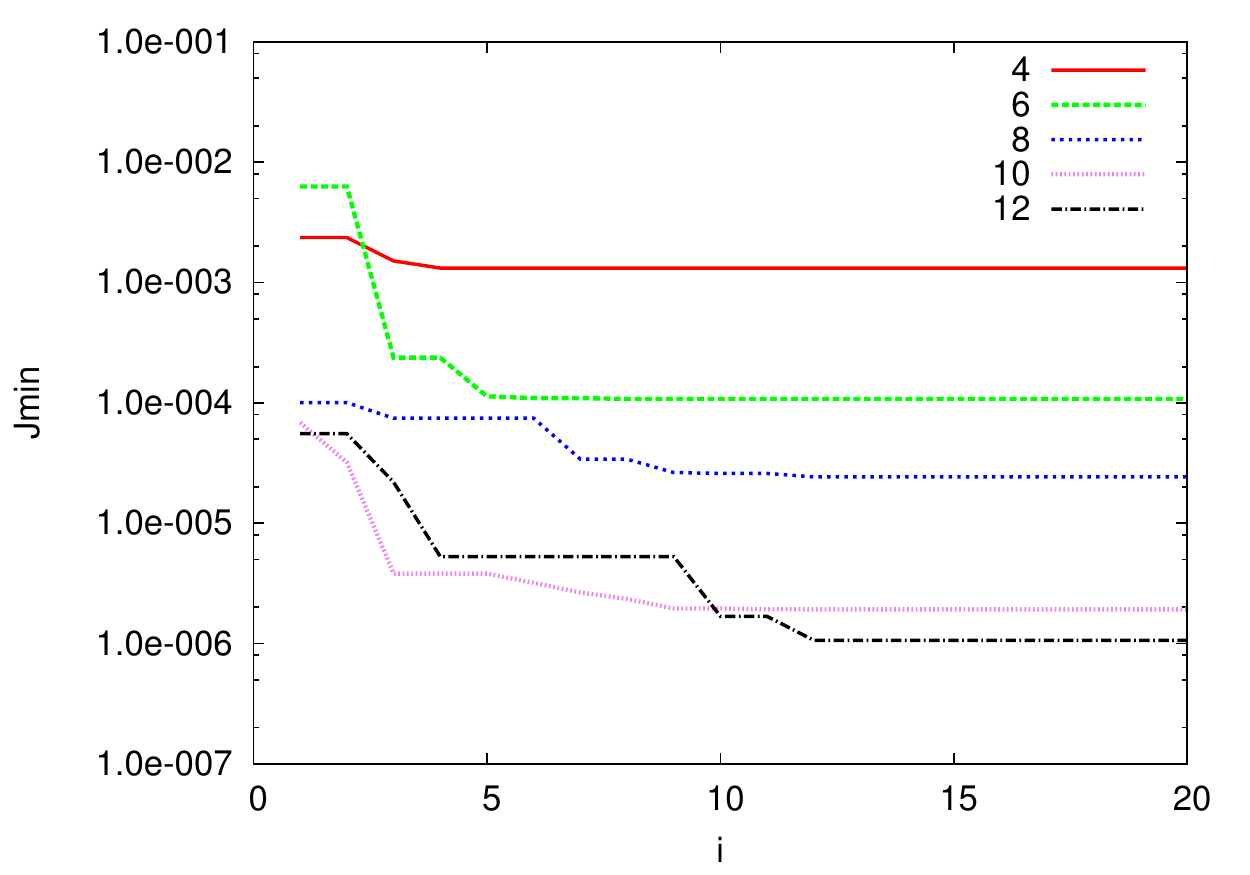} 
\caption{Dependence of the minimum value $J_{min}$ of functional $J$ on the iteration number $i$ for different numbers of
shell layers: $M = 4, 6, 8, 10, 12$ ($k_b = k_0$, $k_{min} = 0.05 \, k_0$, $k_{max} = 236 \, k_0$).}
\end{center}
\end{figure}

\begin{table}[th]
\caption{\label{table3} Numerical results for optimized multilayer shells 
($k_b = k_0$, $k_{min} = 0.05 \, k_0$, $k_{max} = 401 \, k_0$).}
\begin{center}
\begin{tabular}{ccccc}
\hline
$M$ &  $k_M^{opt} / k_0$  & $J^{opt}$   & $J_i^{opt}$   & $J_e^{opt}$  \\
\hline
  2 \rule{0cm}{4mm}  & $401$  & $1.1 \times 10^{-1}$ & $6.1 \times 10^{-3}$ & $2.2 \times 10^{-1}$ \\
  4  & $5.14$  & $7.8 \times 10^{-4}$ & $1.6 \times 10^{-3}$ & $7.8 \times 10^{-6}$ \\
  6  & $5.62$  & $2.8 \times 10^{-5}$ & $5.1 \times 10^{-5}$ & $5.4 \times 10^{-6}$ \\
  8  & $4.31$  & $3.4 \times 10^{-6}$ & $3.6 \times 10^{-6}$ & $3.2 \times 10^{-6}$ \\
  10 & $2.49$  & $8.4 \times 10^{-7}$ & $4.8 \times 10^{-7}$ & $7.1 \times 10^{-7}$ \\
  12 & $0.54$  & $3.5 \times 10^{-7}$ & $8.4 \times 10^{-8}$ & $6.2 \times 10^{-7}$ \\
\hline
\end{tabular}
\end{center}
\end{table}

In our third test we assume that $k_{min} = 0.05 \, k_0$ (wood), 
$k_b = k_0$ (glass) while $k_{max} = 401 \, k_0$ (copper). 
Numerical results for this test are presented in Table~4 (which is analogue of Tables~2 and 3).
As in the previous tests, the optimal values of all control parameters with odd indices coincide and they equal to the lower bound 
$k_{min}=0.05 \, k_0$ while all even controls except the last one coincide with upper bound $k_{max}$.
But it is clearly seen that in this case the optimal values $J^{opt}$ of the functional $J$ 
for all $M$ are substantially less than in Table~2 and in Table~3. 
Thus for the new choice of bound $k_{max}$ we obtained the shell with the highest cloaking efficiency.
This can be explained by the fact that for the third test the value (contrast) $k_{max}/ k_{min}$  is much larger than 
for the first and second tests (8020 versus 400 and 4720). 
Comparing the values of functional $J$ in Tables~2, 3 and 4, we come to the conclusion 
that the cloaking efficiency of the shell increases with increasing number of layers $M$ and the ratio $k_{max}/ k_{min}$.

Since the optimal values of all control parameters except the last one $k_M^{opt}$
are equal to the lower or upper bounds of the control set $K$, the designing a multilayer shell can be reduced to 
solving  one-parameter control problem for $k_M$. 
According to this reduced design,
conductivities of all odd layers are initially prescribed to $k_{min}$,
conductivities of all even layers except $k_M$ are fixed to $k_{max}$ 
and conductivity of the last layer is found by solving 
one-parameter control problem for $k_M$. 
As a result, we can use only three isotropic materials to construct a multilayer shell if 
the optimal conductivity $k_M^{opt}$ corresponds to some natural material.
For example, the value $k_{10}^{opt}=2.49 \, k_0$ from Table~3 at $M=10$ is the conductivity of marble.
If the optimal value $k_M^{opt}$ does not correspond to any natural material than
we can choose materials with conductivities $\tilde k_M^{opt}$  closed to optimal $k_M^{opt}$.
For example, instead of the value $k_{12}^{opt}=8.04 \, k_0$ from the last row of Table~2 we can use 
the close value $\tilde k_{12}^{opt}=7.8  \, k_0$ that corresponds to manganese.
Assuming $\tilde {\bf k}^{opt}=(k_1^{opt},\ldots,k_{M-1}^{opt},\tilde k_M^{opt})$ we obtain 
cloaking shell $(\Omega, \tilde {\bf k}^{opt})$ with $J( \tilde {\bf k}^{opt})=2.8 \times 10^{-4}$
which corresponds to high cloaking efficiency.
Instead of the value $k_{12}^{opt}=0.54 \, k_0$ from the last row of Table~3 we can use 
the value $\tilde k_{12}^{opt}=0.5  \, k_0$ that corresponds to polyethylene.
In this case  $J( \tilde {\bf k}^{opt})=2.49 \times 10^{-4}$.
If the choice of materials is limited but we want to increase cloaking efficiency 
then one can use the geometric parameters of the shell (e.g., the thickness of the last sublayer) as additional controls.

\section{Conclusion}
In this paper we studied inverse problems for 2D model of heat transfer (\ref{eq1})--(\ref{eq6})
associated with designing cylindrical layered thermal cloaking shells. Using the optimization method these inverse problems
were reduced to corresponding control problems in which the thermal conductivities of layers play the role of controls. 
For solving our finite-dimensional extremum problems 
we proposed numerical algorithm based on the particle swarm optimization and discussed simulation results. 
Optimization analysis showed that the optimal values of all control parameters with odd indices 
(conductivities of odd layers) coincide and equal to the lower bound 
of the control set while all even controls except the last one equal to the upper bound of the control set. 
As for the last control, its finding is reduced to solving a one-parameter control problem.
Thus the proposed method simplifies substantially the solution of the cloak design problem under study 
and provides a high cloaking efficiency and a simplicity of technical realization 
of cloaks designed with the help of our method. 
In fact, it is enough to use three different natural materials for obtaining high performance cloaks.
Two of them correspond to the initially specified natural materials with great contrast while 
the latter is found by solving a one-parameter control problem.
Finally, we note that the proposed method and results are not limited only to the temperature field. 
They also could be extended to dc electric, magnetic and other static fields and, besides, to design problems of concentrators, inverters and another functional devices.

\section*{Acknowledgement}
The first author was supported by the Russian Science Foundation (project No. 14-11-00079) and the Russian Foundation for Basic Research (project no. 16-01-00365-a),
the second author was supported by the Federal Agency for Scientific Organizations in the framework of the state task
(subject no. 0263-2018-0001).

\end{document}